\documentclass[12pt]{article}

\usepackage{amsmath}
\usepackage{amssymb}
\usepackage{amsfonts}
\usepackage{latexsym}
\usepackage{bm}
\usepackage{color}

\catcode `\@=11 \@addtoreset{equation}{section}

\catcode `\@=12

\newcommand{\be}{\begin{equation}}
\newcommand{\en}{\end{equation}}
\newcommand{\bea}{\begin{eqnarray}}
\newcommand{\ena}{\end{eqnarray}}
\newcommand{\beano}{\begin{eqnarray*}}
\newcommand{\enano}{\end{eqnarray*}}
\newcommand{\bee}{\begin{enumerate}}
\newcommand{\ene}{\end{enumerate}}

\newcommand{\mc}{\mathcal}

\newcommand{\D}{{\mc D}}
\newcommand{\Dp}{{\D}_{phys}}

\newcommand{\Sc}{{\cal S}}
\newcommand{\E}{{\cal E}}
\newcommand{\F}{{\cal F}}

\newcommand{\Lc}{{\cal L}}

\newcommand{\1}{1 \!\! 1}

\newcommand{\Hil}{\mc H}

\catcode `\@=11 \@addtoreset{equation}{section}
\catcode `\@=12

\begin{document}

\thispagestyle{empty}

\vspace*{2cm}

\begin{center}
{{\Large \bf Transition probabilities for non self-adjoint Hamiltonians in infinite dimensional Hilbert spaces}}\\[10mm]


{\large F. Bagarello} \footnote[1]{ Dipartimento di Energia, Ingegneria dell'Informazione e Modelli Matematici,
Facolt\`a di Ingegneria, Universit\`a di Palermo, I-90128  Palermo, and INFN, Universit\`a di di Torino, ITALY\\
e-mail: fabio.bagarello@unipa.it\,\,\,\, Home page: www.unipa.it/fabio.bagarello}


\end{center}

\vspace*{2cm}

\begin{abstract}
\noindent In a recent paper we have introduced several possible inequivalent descriptions of the dynamics and of the transition probabilities of a quantum system when its Hamiltonian  is not self-adjoint.  Our analysis was carried out in finite dimensional Hilbert spaces.  This is useful, but quite restrictive since many physically relevant quantum systems live in infinite dimensional Hilbert spaces. In this paper we consider this situation, and we discuss some applications to well known models, introduced in the literature in recent years: the extended harmonic oscillator, the Swanson model and a generalized version of the Landau levels Hamiltonian. Not surprisingly we will find new interesting features not previously found in finite dimensional Hilbert spaces, useful for a deeper comprehension of this kind of physical systems.

\end{abstract}


\vfill


\newpage

\section{Introduction}

In ordinary quantum mechanics one of the fundamental axiom of the whole theory is that the Hamiltonian $H$ of the physical system is self-adjoint: $H=H^\dagger$. This condition, shared also by all the { observables} of the system, is important since it ensures that the eigenvalues of these observables, and of the Hamiltonian in particular, are real quantities. Moreover, since the time evolution deduced out of $H$ is unitary, it preserves the total probability: if $\Psi(t)$ is a solution of the Schr\"odinger equation $i\dot\Psi(t)=H\Psi(t)$, then $\|\Psi(t)\|^2$ does not depend on time. This is clear since, if $H$ does not depend explicitly on time, $\Psi(t)=e^{-iHt}\Psi(0)$, and $e^{-iHt}$ is unitary, hence norm-preserving. Of course, this is false if $H\neq H^\dagger$, and in fact, in this case,  $\|\Psi(t)\|^2$ does indeed depend on time, in general. Sometimes this is exactly what one looks for: in many simple systems in quantum optics, for instance, non self-adjoint Hamiltonians are used to describe some decay, so that there is no reason for the probability to be preserved in time. In other situations, one would prefer to avoid any damping, so that the aim is to find some way to {\em recover unitarity} even when $H\neq H^\dagger$, and in fact several attempts have been proposed along the years by different authors to discuss this and other aspects of time evolution for systems driven by non self-adjoint Hamiltonians. Here we refer to \cite{brody}-\cite{mosta3}, and references therein. We also suggest \cite{ben,ali} for two rather general, but not so recent, reviews on this and related subjects, and \cite{book} for a more recent volume, rather mathematically oriented.  Recently, in \cite{bagAoP}, we have discussed some dynamical aspects of this kind of systems, working always with finite-dimensional Hilbert spaces, to avoid dealing with unbounded operators and to  use  the elegant mathematics of pseudo-fermions (PFs), \cite{bagpf,bagpf2,baggar}, in the analysis of the systems. The aim of this paper is to discuss what should be changed and considered with more attention when going from a finite to an infinite dimensional Hilbert space. We will see that the general structure survives to this transition, but that, not surprisingly, several mathematical subtleties must be properly taken into account. Also, we will discuss how this transition produces non trivial physical consequences.

The paper is organized as follows:

In the next section we review the general functional structure associated to a non self-adjoint Hamiltonian, and its dynamics, as well as few inequivalent definitions of transition probability functions, recalling what was done in \cite{bagAoP}. In particular, we discuss what can happens in presence of a metric operator.  In Section \ref{sectexamples} we apply our general results to some well known (and reasonably simple) models, the extended quantum hamonic oscillator, the Swanson model, and a generalized version of the Landau levels, see \cite{bagbook} for a recent review on these models.

Our conclusion is that it may be not so convenient to adopt a different scalar product, the one which makes of $H$ a self-adjoint operator, even if this is possible, in many explicit situations. The reason is that, as we will see in Sections \ref{sectswanmod}  and \ref{sectLL}, this will force us to somehow change the original model, by imposing extra constraints on the coefficients appearing in the definition of the model, coefficients which, in our examples, {\it measure} the non self-adjointness of the Hamiltonian.

\section{The general settings for $H\neq H^\dagger$}

As we have already said, in this paper we will mainly be interested in infinite dimensional Hilbert spaces. Then, our operators can be unbounded, as it very often happens in quantum mechanics, even for very simple systems\footnote{It is well known, in fact, that the Hamiltonian $H$ of one of the simplest, and more important, quantum mechanical system, the harmonic oscillator, is unbounded, as well as the lowering and raising operators in terms of which $H$ can be factorized.}. The main ingredient is an operator $H$, acting on the Hilbert space $\Hil$, with $H\neq H^\dagger$, and with all (multiplicity one) real distinct eigenvalues $E_n$, $n=0,1,2,\ldots$. This condition is useful only to simplify the notation, and could easily be revomed. Here, the adjoint $H^\dagger$ of $H$ is the usual one, i.e. the one defined in terms of the {\em natural} scalar product $\left<.,.\right>$ of the Hilbert space $\Hil$:
$\left<Xf,g\right>=\left<f,X^\dagger g\right>$, for all $f,g\in \Hil$ such that $f\in D(X)$ and $g\in D(X^\dagger)$, the domains of $X$ and $X^\dagger$ respectively. We call $\|.\|$ the norm defined by $\left<.,.\right>$.

In this paper we will consider also two other scalar products on $\Hil$, $\left<.,.\right>_\varphi$ and $\left<.,.\right>_\Psi$. The reason for that will be clarified in the following but can be simply understood already at this level: even if $\left<Hf,g\right>\neq\left<f,H  g\right>$ in general, it will happen that $\left<Hf,g\right>_\Psi=\left<f,H g\right>_\Psi$ and that $\left<H^\dagger \tilde f,\tilde g\right>_\varphi=\left<\tilde f,H^\dagger \tilde g\right>_\varphi$. Of course, due to the fact that $H$ may be unbounded, we have to take $f,g\in D(H)$ and $\tilde f, \tilde g\in D(H^\dagger)$, and these two sets must be, if $H$ and $H^\dagger$ are unbounded, dense subsets of $\Hil$. Needless to say, nothing like this occurs for finite dimensional Hilbert spaces, which is the case most of the time considered in the literature, see for instance \cite{brody,bagAoP}: in fact, in this case, both $D(H)$ and $D(H^\dagger)$ coincide with the whole Hilbert space.

We consider here an Hamiltonian $H$ with distinct real eigenvalues, corresponding to different eigenvectors $\varphi_k\in\Hil$, $k=0,1,2,3,\ldots$:
\be
H\varphi_k=E_k\varphi_k.
\label{21}\en
The set $\F_\varphi=\{\varphi_k,\,k=0,1,2,3\ldots\}$, in many papers, is assumed to be a basis for $\Hil$. This is surely true if $\dim(\Hil)<\infty$, because the various vectors are automatically linearly independent, since they correspond to different eigenvalues. However, in several recent (and not so recent) studies, this has been shown not to be true in general, see \cite{bagbook} and references therein, when $\dim(\Hil)=\infty$. Therefore the assumption that $\F_\varphi$ is a basis for $\Hil$, if $dim(\Hil)=\infty$, is not entirely justified. On the other hand, in all the systems in our knowledge, what is true is that $\F_\varphi$ is a complete set in $\Hil$. This is, of course, a big mathematical difference\footnote{A set $X=\{x_n, \,n\in{\Bbb N}\}$ is complete in $\Hil$ if, for any $f\in\Hil$ and for any $\epsilon>0$, it exists $N>0$ and a set of (complex) coefficients $c_k^{(N)}$, $k=1,2,\ldots,N$, such that $\|f-\sum_{k=1}^Nc_k^{(N)}x_k\|\leq\epsilon$. In general these coefficients depend on $N$. When it happens that they {\bf do not} depend on $N$, the set $X$ is a basis for $\Hil$ since $\|f-\sum_{k=1}^Nc_kx_k\|\rightarrow0$ for $N$ diverging. It is well known that completeness of $X$ is equivalent to $X$ being a basis if the vectors $x_n$'s are mutually orthonormal. Otherwise this is not true: any basis (orthonormal or not) is complete, but not all complete sets are bases.}, \cite{heil}. For this reason, we will restrict here to this lighter, and always satisfied, assumption: $\F_\varphi$ is complete  in $\Hil$. To this working assumptions we have to add similar conditions concerning $H^\dagger$ which, as an operator on $\Hil$, may appear very different from $H$. In particular we will assume here that $H^\dagger$ admits eigenvectors $\Psi_k$ with the same eigenvalues as $H$:
\be
H^\dagger\Psi_k=E_k\Psi_k.
\label{22}\en
Hence $H$ and $H^\dagger$ are assumed to be isospectrals, in this paper. The set $\F_\Psi=\{\Psi_k,\,k=0,1,2,3\ldots\}$ will also be assumed to be complete in $\Hil$, but not necessarily a basis. Of course,  together with $\F_\varphi$, these are biorthogonal sets: $\left<\varphi_n,\Psi_k\right>=\delta_{n,k}$.

\vspace{2mm}

Similarly to what is proposed in \cite{ali2}, rather than working on $\Hil$ with the complete sets $\F_\varphi$ and $\F_\Psi$, it is convenient to define a subset of $\Hil$, which we call $\Dp$, as follows:
\be
\Dp:=\left\{f\in\Hil:\, f=\sum_n\left<\varphi_n,f\right>\Psi_n=\sum_n\left<\Psi_n,f\right>\varphi_n\right\}.
\label{23}\en
Then, $\Dp$ is not necessarily all of $\Hil$, except, for instance, if $\Hil$ is finite dimensional or if $\F_\varphi$ and $\F_\Psi$ are biorthogonal (Riesz) bases. The subscript {\em phys} stands for {\em physical}, meaning with this that the set $\Dp$ is assumed to contain all the physically relevant vectors needed for the full description of the physical system $\Sc$ we are interested in. In other words, even if $\F_\varphi$ and $\F_\Psi$ are not bases for $\Hil$, they are still sufficient to expand all those vectors which have a physical meaning. Of course, an obvious requirement is that each $\varphi_n$ and each $\Psi_n$ belong to $\Dp$, since these vectors are surely the {\em most physical} of the system, being eigenstates of $H$ and $H^\dagger$. Due to the biorthogonality of $\F_\varphi$ and $\F_\Psi$ this is equivalent to require that, for all $k$,
\be
\varphi_k=\sum_n\left<\varphi_n,\varphi_k\right>\Psi_n, \qquad \Psi_k=\sum_n\left<\Psi_n,\Psi_k\right>\varphi_n.
 \label{24}\en
Of course, these equalities surely hold whenever  $\F_\varphi$ and $\F_\Psi$ collapse to a single orthonormal (o.n.) basis of $\Hil$,  when they are Riesz bases, or for finite-dimensional Hilbert spaces. In other cases, however, they must be explicitly checked. Notice that, since $\F_\varphi$ and $\F_\Psi$ are complete in $\Hil$, $\Dp$ is dense in $\Hil$. Then, $\Dp$ is a large set, indeed.

\vspace{2mm}

{\bf Remark:--} It could be interesting to notice that, while here the physical space is defined by $\F_\varphi$ and $\F_\Psi$, which in turns are determined by $H$ and $H^\dagger$, in \cite{ali2} the Hilbert space where the model lives has to do with the set of all the observables of the system. This is a natural procedure, of course, even if it is not guaranteed that the intersection of all these domains is a {\em sufficiently large} set, in concrete situations. We also would like to notice that, contrarily to what stated in \cite{ali2}, the analysis carried out in this paper will suggest that the natural scalar product to be considered in the physical space is exactly the one originally defined in $\Hil$. This will be clarified in the second part of the paper.

\vspace{2mm}

In \cite{bagAoP} we have discussed how the dynamics of a physical system $\Sc$ should be defined when its Hamiltonian $H$ is not self-adjoint. Our suggestion, which agrees with the point of view widely adopted in the literature, is that the wave function $\Phi(t)$ of $\Sc$ should satisfy a standard Schr\"odinger equation, $i\dot \Phi(t)=H\Phi(t)$, with $\Phi(0)=\Phi_0$. Of course, it is natural to require first that $\Phi_0\in\Dp$, and to assume also that this property is preserved under time evolution. In other words, we would like to have $\Phi(t)\in\Dp$ for all $t\geq0$, and not just for $t=0$. This is not granted, because it is not necessarily true that, even if $\Phi_0=\sum_n\left<\varphi_n,\Phi_0\right>\Psi_n=\sum_n\left<\Psi_n,\Phi_0\right>\varphi_n$, then
\be
\Phi(t)=\sum_n\left<\varphi_n,\Phi(t)\right>\Psi_n=\sum_n\left<\Psi_n,\Phi(t)\right>\varphi_n.
\label{25}\en
However, this is the case if, for instance, $(i)$ $e^{-iHt}$ is bounded and $(ii)$ $\sum_n\left<\varphi_n,\Phi(t)\right>\Psi_n$ is a Cauchy sequence\footnote{Of course, this is surely true for $t=0$. We are here assuming that this is also true for $t>0$.} in $\Hil$. In fact, under these assumptions we have:
$$
\Phi(t)=e^{-iHt}\Phi_0=e^{-iHt}\left(\sum_n\left<\Psi_n,\Phi_0\right>\varphi_n\right)=\sum_n\left<\Psi_n,\Phi_0\right>e^{-iHt}\varphi_n,
$$
using the continuity of $e^{-iHt}$. Hence
$$
\Phi(t)=\sum_n\left<\Psi_n,\Phi_0\right>e^{-iE_nt}\varphi_n=\sum_n\left<e^{iE_nt}\Psi_n,\Phi_0\right>\varphi_n=
\sum_n\left<e^{iH^\dagger t}\Psi_n,\Phi_0\right>\varphi_n=
$$
$$
=\sum_n\left<\Psi_n,e^{-iH t}\Phi_0\right>\varphi_n=\sum_n\left<\Psi_n,\Phi(t)\right>\varphi_n,
$$
which is half of equation (\ref{25}). As for the second half, since $\sum_n\left<\varphi_n,\Phi(t)\right>\Psi_n$ is, by assumption, a Cauchy sequence for all $t\geq0$, it surely converges to some vector of $\Hil$. The fact that this vector is exactly $\Phi(t)$ follows from the fact that $\F_\varphi$ is complete in $\Hil$ and by the biorthogonality of $\F_\varphi$ and $\F_\Psi$, since we have $\left<\varphi_k,\left[\Phi(t)-\sum_n\left<\varphi_n,\Phi(t)\right>\Psi_n\right]\right>=0$ for all $k$.

\vspace{1mm}

This result shows that, when the dynamics of the physical system $\Sc$ is driven by a non self-adjoint Hamiltonian $H$, problems arise both at a pure algebraic level and at the level of its dynamical description. If, from one side, the introduction of $\Dp$ looks quite reasonable on a physical ground, requiring that $\Dp$ is stable under time evolution is not granted a priori, and some extra conditions are required. Of course, the ones given here are sufficient conditions, so we can imagine that they can be lightened.   Once again, all the conditions are satisfied if $\dim(\Hil)<\infty$, which is the situation quite often discussed in the literature, or when $\F_\varphi$ and $\F_\Psi$ are biorthogonal Riesz bases.

\vspace{2mm}

{\bf Remarks:--} (1) Of course, these steps can be slightly modified, with few and obvious changes, if one assumes as driving Hamiltonian $H^\dagger$ rather than $H$.

\vspace{1mm}

(2) The dynamics of the operators in the Heisenberg, as well as in the Schr\"odinger, representation  is not uniquely defined. A natural choice is
\be
 X (t)=e^{iH^\dagger t} Xe^{-iHt}.
\label{26}\en
As we have discussed in \cite{bagAoP}, this is  not the only possibility, and it is not necessarily the most convenient, since neither $e^{iH^\dagger t}$ nor $e^{-iH t}$ are unitary (which however is exactly what one looks for, sometimes). Moreover, adopting the rule in (\ref{26}) it is not so easy to find integrals of motion for the system, since $[H,\hat X]=0$ does not imply that $  X(t)=  X(0)$ for all $t\geq0$. Finally, a serious difficulty is that the time evolution is no longer an automorphism of the set of observables, since in general $(\hat X\hat Y)(t)\neq \hat X(t)\hat Y(t)$, and this complicates in an enormous way all the computations. We refer to \cite{bagAoP} for further considerations on this aspect, which is not our main concern here.

\vspace{1mm}

(3) It might be more convenient to replace the set $\Dp$ with $$ \Dp^w:=\left\{f\in\Hil:\, f=\sum_n\left<\varphi_n,f\right>\Psi_n, \quad\mbox{or}\quad f=\sum_n\left<\Psi_n,f\right>\varphi_n\right\}.$$
Of course, each element of $\Dp$ also belongs to $\Dp^w$, while the vice-versa is not true. This means, in particular, that $\Dp^w$ is dense in $\Hil$. We prefer $\Dp$ because of a more evident symmetry between $\F_\varphi$ and $\F_\Psi$.

\vspace{2mm}

As in \cite{bagAoP}, we are mainly interested here in defining the probability transition between two states, the initial state of the physical system $\Sc$, $\Phi_0$, and the final state, $\Phi_f$. As we have discussed in \cite{bagAoP}, this definition is not unique. This is due to the presence of, at least, three inequivalent scalar products defined in $\Dp$ or even in $\Hil$. In analogy with \cite{bagAoP}, we call these products  $\left<.,.\right>$, $\left<.,.\right>_\Psi$ and $\left<.,.\right>_\varphi$, and we call $\flat$ and $\sharp$ the related adjoint maps: $\left<Xf,g\right>_\Psi=\left<f,X^\sharp g\right>_\Psi$ and $\left<Xf,g\right>_\varphi=\left<f,X^\flat g\right>_\varphi$, for all $f$ and $g$ in $\Hil$ and for all operators $X$ for which these equalities make sense. The products $\left<.,.\right>_\Psi$ and $\left<.,.\right>_\varphi$ are such that, as discussed before, $H^\dagger$ and $H$ are self-adjoint with respect to them:
\be
\left<f,Hg\right>_\Psi=\left<Hf,g\right>_\Psi, \qquad \left<H^\dagger \tilde f,\tilde g\right>_\varphi=\left<\tilde f,H^\dagger \tilde g\right>_\varphi,
\label{26b}
\en
for all $f,g\in D(H)$ and $\tilde f,\tilde g\in D(H^\dagger)$. Here we are implicitly assuming that both $D(H)$ and $D(H^\dagger)$ are subsets of $\Dp$. Using (\ref{26b}) we deduce that $H=H^\sharp$, and that $H^\dagger=(H^\dagger)^\flat$. Once we have three scalar products, we also have several different possible definitions of the transition probabilities. The ones we consider here are the following:

\be
P_{\Phi_0\rightarrow\Phi_f}(t):=\left|\frac{\left<\Phi_f,\Phi(t)\right>}{\|\Phi_f\|\|\Phi(t)\|}\right|^2, \, P_{\Phi_0\rightarrow\Phi_f}^\Psi(t):=\left|\frac{\left<\Phi_f,\Phi(t)\right>_\Psi}{\|\Phi_f\|_\Psi\|\Phi(t)\|_\Psi}\right|^2, \, P_{\Phi_0\rightarrow\Phi_f}^\varphi(t):=\left|\frac{\left<\Phi_f,\Phi(t)\right>_\varphi}{\|\Phi_f\|_\varphi\|\Phi(t)\|_\varphi}\right|^2.
\label{27}\en
These were already introduced and analyzed in \cite{bagAoP}, for systems living in finite dimensional Hilbert spaces, and we have seen that they produce different results, so that, in fact, \underline{they are not} \underline{physically equivalent at all}. Then we have proposed to discriminate among these three definitions using some concrete experiment, and we have considered a simple two-level system. Here we want to carry on a similar analysis, but considering systems which live in an infinite dimensional Hilbert space. This will be done in Section \ref{sectexamples}, where we will see how it is possible, in principle, to discriminate among the functions in (\ref{27}), in order to understand which is the most appropriate expression of the transition probability, and why. Incidentally we observe that, with these definitions, the images of $P_{\Phi_0\rightarrow\Phi_f}(t)$, $P_{\Phi_0\rightarrow\Phi_f}^\Psi(t)$ and $P_{\Phi_0\rightarrow\Phi_f}^\varphi$(t) is always the set $[0,1]$, for all $t\geq0$.

\vspace{2mm}

{\bf Remark:--} If we were interested in keeping the time evolution unitary, then $P_{\Phi_0\rightarrow\Phi_f}^\Psi$ would be the more natural choice, since $H$ is self-adjoint with respect to $\left<.,.\right>_\Psi$, and therefore $e^{-iHt}$ is unitary with respect to this scalar product. However, here we  are much more interested in a comparison between the theoretical results with some experimental data. This is also relevant in view of the fact that,  if we replace $H$ with the, equally valid, operator $H^\dagger$, the unitarity requirement would suggest, of course, to use  $P_{\Phi_0\rightarrow\Phi_f}^\varphi$. In other words: different ingredients produce different rules.

\vspace{2mm}

Interestingly enough, we will see that the analysis of the functions in (\ref{27}) in some concrete models suggests to avoid the use of $\left<.,.\right>_\varphi$ and $\left<.,.\right>_\Psi$, and to restrict to $\left<.,.\right>$, and to $P_{\Phi_0\rightarrow\Phi_f}(t)$ as a consequence. The other choices, in fact, sometimes produce a unwanted extra constraint in the range of the parameters of the model, in order to make sense out of the model itself. This will be made explicit in Sections \ref{sectswanmod} and \ref{sectLL}.

\subsection{Refining the structure}

What we have discussed so far does not imply the existence of any particular relation between the three scalar products above. They are just related, in principle, to $H$ and $H^\dagger$. This is because the two {\em new} scalar products are introduced here just to make of $H$ and $H^\dagger$ two self-adjoint operators.
In many physical systems considered in the literature, however, a relation between them does in fact exist, and it is provided by the so-called {\em metric operator}. Again, while there is no problem to introduce this operator if $\dim(\Hil)<\infty$, serious problems may occur for infinite-dimensional Hilbert spaces. The reason is that it may happen that this operator, or its inverse, or both, are unbounded. When this happens, we have to pay attention to domains. In particular, we define
$$
D(S_\varphi):=\left\{f\in\Hil:\, \sum_n\left<\varphi_n,f\right>\varphi_n\in\Hil\right\}, \qquad D(S_\Psi):=\left\{g\in\Hil:\, \sum_n\left<\Psi_n,g\right>\Psi_n\in\Hil\right\},
$$
and
\be
S_\varphi f=\sum_n\left<\varphi_n,f\right>\varphi_n. \qquad S_\Psi g=\sum_n\left<\Psi_n,g\right>\Psi_n,
\label{28}\en
 for all $f\in D(S_\varphi)$ and $g\in D(S_\Psi)$. To make the situation technically simpler, it is convenient to work  under the assumption that $\Dp\subseteq D(S_\varphi)\cap D(S_\Psi)$. This makes of $S_\varphi$ and $S_\Psi$ two densely defined operators, if $\Dp$ is dense in $\Hil$, as we have observed in several concrete examples discussed so far, \cite{bagbook}. These operators have the following properties: (i) $\Psi_n\in D(S_\varphi)$, and $S_\varphi\Psi_n=\varphi_n$; (i) $\varphi_n\in D(S_\Psi)$, and $S_\Psi\varphi_n=\Psi_n$; (iii) they are positive operators, and, under suitable conditions, they admit a self-adjoint (Friedrichs) extension, which are also positive, and which we indicate with the same symbols; (iv) these extensions admit square roots, $S_\varphi^{1/2}$ and $S_\Psi^{1/2}$; (v) for all $f,g\in\Dp$ the scalar products introduced above are related as follows:
 $$
 \left<f,g\right>_\Psi=\left<f,S_\Psi g\right>=\left<S_\Psi^{1/2}f,S_\Psi^{1/2}g\right>,
 $$
and
$$
 \left<f,g\right>_\varphi=\left<f,S_\varphi g\right>=\left<S_\varphi^{1/2}f,S_\varphi^{1/2}g\right>.
 $$
Another feature of $S_\varphi$ and $S_\Psi$ is that, again under suitable assumptions, they relate the different adjoint introduced so far, $\dagger$, $\flat$ and $\sharp$. In fact, taken an operator $X$ of $\Sc$, and assuming for simplicity that $X$ leaves invariant $\Dp$ together with $X^\dagger$, $X^\flat$ and $X^\sharp$, we deduce the following equalities:
$$
X^\flat f=S_\Psi X^\dagger S_\varphi f,\qquad X^\sharp f=S_\varphi X^\dagger S_\Psi f\qquad \mbox{and}\quad X^\flat f=S_\Psi^2 X^\sharp S_\varphi^2 f,
$$
for all $f\in\Dp$. Of course, we are also assuming that $S_\varphi$ and $S_\Psi$ leave $\Dp$ invariant. Finally a direct computation shows that
$$
 S_\Psi H \varphi_n=H^\dagger S_\Psi\varphi_n,\qquad S_\varphi H^\dagger\Psi_n= HS_\varphi\Psi_n.
$$

\vspace{2mm}

{\bf Remark:--}  The results sketched in this section suggests that it is the Hamiltonian $H$ itself which somehow fixes its preferred Hilbert space. This is because both $\left<.,.\right>_\varphi$ and $\left<.,.\right>_\Psi$ are defined via $S_\varphi$ and $S_\Psi$, which are constructed, in turns, by the eigenvectors of $H$ and $H^\dagger$. This is similar to what happens in algebraic quantum dynamics, see \cite{bagrev} and references therein, where the Hamiltonian (self-adjoint, in that context) is used to define a suitable topology on the algebra of the operators needed in the description of the physical system. This aspect is also discussed in \cite{bagAoP}, together with the role of non zero temperature states.

\vspace{2mm}

If we now introduce formally $H_0:=S_\Psi^{1/2}HS_\varphi^{1/2}$ and $e_k=S_\Psi^{1/2}\varphi_k= S_\varphi^{1/2}\Psi_k$,  we see that $H_0=H_0^\dagger=S_\varphi^{1/2}H^\dagger S_\Psi^{1/2}$,  and that $\E=\{e_k,\,k=0,1,2,\ldots,\}$ is an orthonormal (o.n.) set of $\Hil$ of eigenstates of $H_0$: $H_0e_k=E_ke_k$. To move from formal to rigorous results we need to perform, of course, a deeper analysis of the operators considered. In particular, if for some reason $\Dp$ is left invariant by $H$, $S_\Psi^{1/2}$ and $S_\varphi^{1/2}$, as it happens in some concrete examples, \cite{bagbook},  then $H_0$ turns out to be a densely defined symmetric operator and, if $\F_\varphi$ is a Riesz basis, then $\E$ is an o.n. basis.

\vspace{2mm}

{\bf Remark:--} A possible alternative approach consists in introducing, together with $\Dp$, the set of what we can call {\em physically relevant operators}, ${\cal O}_{phys}$, as the set of all the operators $X$, bounded or not, densely defined on $\Hil$, which leave stable $\Dp$ together with all their adjoints, $X^\dagger$, $X^\sharp$ and $X^\flat$. Then, our working assumption is that $H$, $e^{iHt}$,  $S_\varphi^{1/2}$ and  $S_\Psi^{1/2}$ belong to ${\cal O}_{phys}$. This is not very different from what it is done in the literature on unbounded operator algebras when one introduces the set
$\Lc^\dagger(\D)$, which is the *-algebra of all the {  closable operators}
defined on the dense set $\D$ which, together with their adjoints, map $\D$ into
itself, \cite{bagrev}. In fact, also in our case, the set ${\cal O}_{phys}$ turns out to be an algebra of unbounded operators, having $\Dp$ as the common domain. One of the obvious differences between ${\cal O}_{phys}$ and $\Lc^\dagger(\D)$ is that the first one involves several involutions, while the latter just one\footnote{This is not really so, because one has also to deal with the restriction of the involution to $\D$, but this has nothing to do what we are considering in this paper.}.

\section{Examples}\label{sectexamples}

In this section we will consider some concrete examples, already considered in the literature, to compare the different definitions of transition probabilities introduced in (\ref{27}) and to deduce in this way which one, among the different possibilities, is the more appropriate. More concretely, we will analyze what happens for the extended quantum harmonic oscillator (EQHO), for the Swanson model, and for the extended Landau levels (ELLs), see \cite{dapro,bagpbexa,abg}, computing the different expressions of the transition probabilities for particular choices of the initial and the final states. With respect to what has been discussed in \cite{bagAoP}, the role of unbounded operators will clearly show up, and we will see that new phenomena will occur exactly because of the infinite dimensionality of the Hilbert space, leading to conclusions somewhat different from those deduced in \cite{bagAoP}.

\subsection{The extended quantum harmonic oscillator}\label{secteqho}

The Hamiltonian of the EQHO, as originally proposed in \cite{dapro} and then rewritten in terms of pseudo-bosonic operators in \cite{bagpbexa}, looks as follows:
$$H_\nu=\frac{\nu}{2}\left(p^2+x^2\right)+i\sqrt{2}\,p,$$ where $\nu$
is a strictly positive parameter and $[x,p]=i\1$. Here $\1$ is the identity operator on $\Hil=\Lc^2(\Bbb R)$. Of course, $H_\nu$ is manifestly non hermitian.

Introducing the standard bosonic operators $a=\frac{1}{\sqrt{2}}\left(x+\frac{d}{dx}\right)$,
$a^\dagger=\frac{1}{\sqrt{2}}\left(x-\frac{d}{dx}\right)$, $[a,a^\dagger]=\1$, and the related operators $ A_\nu=a-\frac{1}{\nu}$, and
$B_\nu=a^\dagger+\frac{1}{\nu}$, we can write $ H_\nu=\nu\left(B_\nu A_\nu+\gamma_\nu\,\1\right), $ where $\gamma_\nu=\frac{2+\nu^2}{2\nu^2}$. It is
clear that, for all $\nu>0$, $A_\nu^\dagger\neq B_\nu$ and  $[A_\nu, B_\nu]=\1$. Hence we have to do, at least formally, with pseudo-bosonic operators. We refer to \cite{bagbook,dapro,bagpbexa}  for more details. In particular, we have proven that the sets $\F_\varphi=\{\varphi_n^{(\nu)}(x)\}$ and $\F_\Psi=\{\Psi_n^{(\nu)}(x)\}$ of eigenstates of $H_\nu$ and
$H_\nu^\dagger$ are not biorthogonal  bases, but still they are $\D$-quasi bases\footnote{This means that $\F_\varphi$ and $\F_\Psi$ still resolve the identity, but only weakly on the dense set $\D$.}, and that they are both complete in $\Lc^2(\Bbb R)$.

Also, we have deduced that, with a proper choice of normalization,
$$
\varphi_n^{(\nu)}(x)=\frac{e^{-1/\nu^2}}{\pi^{1/4}\,\sqrt{2^n\,n!}}\,\left(x-\frac{d}{dx}+\frac{\sqrt{2}}{\nu}\right)^n\,e^{-\frac{1}{2}(x-\sqrt{2}/\nu)^2},
$$
and
$$
\Psi_n^{(\nu)}(x)=\frac{e^{1/\nu^2}}{\pi^{1/4}\,\sqrt{2^n\,n!}}\,\left(x-\frac{d}{dx}-\frac{\sqrt{2}}{\nu}\right)^n\,
e^{-\frac{1}{2}(x+\sqrt{2}/\nu)^2}.
$$
They both correspond to the same eigenvalue, $E_n^{(\nu)}=\nu(n+\gamma_\nu)$, for $H_\nu$ and $H_\nu^\dagger$ respectively. The so-called metric operator $\Theta_\nu$, mapping $\F_\varphi$ into $\F_\Psi$, is a simple multiplication operator, which looks like
$
\Theta_\nu=e^{2/\nu^2}\,e^{-\frac{2}{\nu}(a+a^\dagger)}=e^{2/\nu^2}\,e^{-2\sqrt{2}\,\frac{x}{\nu}}.
$
We have $\Psi_n^{(\nu)}(x)=\Theta_\nu\varphi_n^{(\nu)}(x)$, for all $n$. It is clear that $\Theta_\nu$ is unbounded, since it is not everywhere defined in $\Lc^2(\Bbb R)$. However, it is invertible with unbounded inverse and we obviously have that $\varphi_n^{(\nu)}(x)\in D(\Theta_\nu)$, and that $\Psi_n^{(\nu)}(x)\in D(\Theta_\nu^{-1})$ for all $n$. Of course, the operator $\Theta_\nu$ must be identified with the operators $S_\Psi$ introduced in (\ref{28}).

In order to compare the transition probabilities in (\ref{27}), and to make the computations simpler, we restrict to the first two eigenstates of $H_\nu$ and $H_\nu^\dagger$:
$$
\varphi_0^{(\nu)}(x)=\frac{e^{-1/\nu^2}}{\pi^{1/4}}\,e^{-\frac{1}{2}(x-\sqrt{2}/\nu)^2}, \qquad \varphi_1^{(\nu)}(x)=\frac{e^{-1/\nu^2}}{\pi^{1/4}}\,\sqrt{2}\,x\,e^{-\frac{1}{2}(x-\sqrt{2}/\nu)^2},
$$
and
$$
\Psi_n^{(\nu)}(x)=\frac{e^{1/\nu^2}}{\pi^{1/4}}\,e^{-\frac{1}{2}(x+\sqrt{2}/\nu)^2}, \qquad
\Psi_1^{(\nu)}(x)=\frac{e^{1/\nu^2}}{\pi^{1/4}}\,\sqrt{2}\,x\,e^{-\frac{1}{2}(x+\sqrt{2}/\nu)^2}.
$$
It is now an easy exercise to compute the transition probabilities for some different choices of $\Phi_0$ and $\Phi_f$. For instance, if $\Phi_0=\varphi_0^{(\nu)}$ and $\Phi_f=\Psi_0^{(\nu)}$, we find that

\be
P_{\Phi_0\rightarrow\Phi_f}(t)=1, \quad P_{\Phi_0\rightarrow\Phi_f}^\Psi(t)=e^{-6/\nu^2}, \quad P_{\Phi_0\rightarrow\Phi_f}^\varphi(t)=e^{-2/\nu^2},
\label{31}\en
which are all different and independent on time. Notice that when $\nu$ is taken large enough the three probabilities all converge to one. This appears in agreement with the fact that, in this limit, the self-adjoint part of $H_\nu$ is much larger than the remaining part, so that, essentially, the deviation form a standard situation is really small.

If we now take $\Phi_0=\Phi_f=\varphi_0^{(\nu)}$ it is easy to see that $
P_{\Phi_0\rightarrow\Phi_f}(t)= P_{\Phi_0\rightarrow\Phi_f}^\Psi(t)= P_{\Phi_0\rightarrow\Phi_f}^\varphi(t)=1$, which is not surprising. In fact, this result can be easily generalized to the following situation: suppose that $\Sc$ is a physical system with (non self-adjoint) Hamiltonian $H$ and let $\varphi_\alpha$ be an eigenstate of $H$. If we choose $\Phi_0=\Phi_f=\varphi_\alpha$, then $
P_{\Phi_0\rightarrow\Phi_f}(t)= P_{\Phi_0\rightarrow\Phi_f}^\Psi(t)= P_{\Phi_0\rightarrow\Phi_f}^\varphi(t)=1$. In this case the three probabilities coincide. Hence, this choice is not useful to discriminate among the various functions in (\ref{27}).

More interesting for us is the situation in which $\Phi_0=\varphi_0^{(\nu)}+\varphi_1^{(\nu)}$ and $\Phi_f=\Psi_0^{(\nu)}$. In this case, obviously, we have $\Phi(t)=e^{-iHt}\Phi_0=e^{-iE_0t}\varphi_0^{(\nu)}+e^{-iE_1t}\varphi_1^{(\nu)}$, and after some computations we conclude that
$$
P_{\Phi_0\rightarrow\Phi_f}(t)=\frac{\nu^2}{2\left(2+\nu^2+2\nu\cos(\nu t)\right)},
$$
$$
P_{\Phi_0\rightarrow\Phi_f}^\Psi(t)=\frac{1}{2}e^{-6/\nu^2}\left(\left(1-\frac{2}{\nu}\right)^2+\frac{4}{\nu}\left(1-\cos(\nu t) \right)\right),
$$
and
$$
P_{\Phi_0\rightarrow\Phi_f}^\varphi(t)=\frac{\nu^2}{2}e^{-6/\nu^2}\frac{\left(1-\frac{2}{\nu}\right)^2+\frac{4}{\nu}\left(1+\cos(\nu t) \right)}{8+\nu^2+4\nu\cos(\nu t)}.
$$
These formulas show that each probability transition go to $\frac{1}{2}$ when $\nu\rightarrow\infty$, for all $t$. This is in agreement with our previous interpretation of this limit. Also, if $\nu\rightarrow0$ then all these functions converge to zero: the $\Phi_0\rightarrow\Phi_f$ transition is not allowed, in this case, whichever choice we do. Notice also that the three functions are periodic, with a period which is exactly $\frac{2\pi}{\nu}$: the smaller the value of $\nu$, the longer the period.




It is not difficult now to imagine, at least in principle, concrete experiments capable to discriminate among the three definitions in (\ref{27}), just comparing the experimental results with what we have deduced above with the first or the third choice of $\Phi_0$ and $\Phi_f$. The second choice is not useful for us, since the resulting functions do coincide. We notice that, in the analysis of the EQHO, there is no  reason to prefer $P_{\Phi_0\rightarrow\Phi_f}(t)$ to the other two possibilities, expect for its agreement with experiments. We will see that this is not what happens for the other models we are going to consider next.

\subsection{The Swanson model}\label{sectswanmod}

It is interesting to discuss also what happens for the Swanson model, since, as we will see in a moment, these are new facts which were not observed in \cite{bagAoP} and in the previous example, the EQHO. Again we refer to \cite{bagbook} for the details of our construction.

The  non self-adjoint Hamiltonian of the model is $$
H_\theta=\frac{1}{2}\left(p^2+x^2\right)-\frac{i}{2}\,\tan(2\theta)\left(p^2-x^2\right), $$ where $\theta$ is a real parameter taking value in
$\left(-\frac{\pi}{4},\frac{\pi}{4}\right)\setminus\{0\}=:I$, \cite{dapro}, and $[x,p]=i\1$.

Introducing now the (standard bosonic) annihilation and
creation operators $a$, $a^\dagger$, and their linear combinations
$$\left\{
\begin{array}{ll}A_\theta=\cos(\theta)\,a+i\sin(\theta)\,a^\dagger=\frac{1}{\sqrt{2}}\left(e^{i\theta}x+e^{-i\theta}\,\frac{d}{dx}\right),\\
B_\theta=\cos(\theta)\,a^\dagger+i\sin(\theta)\,a=\frac{1}{\sqrt{2}}\left(e^{i\theta}x-e^{-i\theta}\,\frac{d}{dx}\right),\end{array} \right. $$
we can write $ H_\theta=\omega_\theta\left(B_\theta\,A_\theta+\frac{1}{2}\1\right), $ where $\omega_\theta=\frac{1}{\cos(2\theta)}$ is well
defined because $\cos(2\theta)\neq0$ for all $\theta\in I$.  The eigenfunctions  of $H_\theta$ and $H_\theta^\dagger$, forming the sets $\F_\varphi^{(\theta)}$ and $\F_\Psi^{(\theta)}$, have been found in \cite{bagpbexa}:
$$
\left\{
\begin{array}{ll}
\varphi_n^{(\theta)}(x)=\frac{e^{i\theta/2}}{\pi^{1/4}\sqrt{2^n\,n!}}
\,H_n\left(e^{i\theta}x\right)\,\exp\left\{-\frac{1}{2}\,e^{2i\theta}\,x^2\right\},  \\
\Psi_n^{(\theta)}(x)=\frac{e^{-i\theta/2}}{\pi^{1/4}\sqrt{2^n\,n!}}
\,H_n\left(e^{-i\theta}x\right)\,\exp\left\{-\frac{1}{2}\,e^{-2i\theta}\,x^2\right\},
\end{array}
\right.
$$
where $H_n(x)$ is the n-th Hermite polynomial. These functions all belong to $\Lc^2({\Bbb R})$, but they are not bases of this space, thought being complete, \cite{bagbook}. The operator $S_\Psi$ which maps $\F_\varphi^{(\theta)}$ into $\F_\Psi^{(\theta)}$ is the following:
$$
\left(S_\Psi f\right)(x)=e^{-i\theta/2}f\left(e^{-i\theta}x\right),
$$
which is, of course, not everywhere defined. Hence,  $S_\Psi$ is unbounded, with unbounded inverse $S_\varphi$ satisfying the following: $\left(S_\varphi f\right)(x)=e^{i\theta/2}f\left(e^{i\theta}x\right)$. It is easy to check that $\varphi_n^{(\theta)}(x)\in D(S_\Psi)$ and that $S_\Psi\varphi_n^{(\theta)}(x)=\Psi_n^{(\theta)}(x)$, for all $n\geq0$. Analogously, we can check that $\Psi_n^{(\theta)}(x)\in D(S_\varphi)$ and that $S_\varphi\Psi_n^{(\theta)}(x)=\varphi_n^{(\theta)}(x)$.

In the attempt to deduce some transition probability function explicitly dependent on time we consider first the initial state of $\Sc$ to be $\Phi_0=\varphi_0^{(\theta)}+\varphi_1^{(\theta)}$, and $\Phi_f=\Psi_0^{(\theta)}$, as in Section \ref{secteqho}. However, this is not enough, due to the fact that the vectors in $\F_\varphi^{(\theta)}$ and $\F_\Psi^{(\theta)}$ are just rotated versions of the eigenstates of a quantum harmonic oscillator, with a slightly different normalization. In fact we get
$$
P_{\Phi_0\rightarrow\Phi_f}(t)=\frac{\cos^2(2\theta)}{1+\cos^2(2\theta)},
$$
which does not depend explicitly on time. A similar result can be deduced also for the other functions in (\ref{27}):
$$
P_{\Phi_0\rightarrow\Phi_f}^\Psi(t)=\frac{\sqrt{\cos(4\theta)}}{2\cos{2\theta}}, \qquad P_{\Phi_0\rightarrow\Phi_f}^\varphi(t)=\frac{(\cos(4\theta))^{3/2}}{\cos{2\theta}(\cos(4\theta)+1)}.
$$
It is important to notice that these two last functions are not defined for all $\theta\in I$, but only if $\theta\in\left(-\frac{\pi}{8},\frac{\pi}{8}\right)\setminus\{0\}=:I_1$. This suggests that the definitions $P_{\Phi_0\rightarrow\Phi_f}^\Psi(t)$ and $P_{\Phi_0\rightarrow\Phi_f}^\varphi(t)$ are somehow artificial, and only make sense if we are willing to change the original model by imposing, as we have to do here, more constraints on the parameters of the system.

\vspace{2mm}

{\bf Remark:--} It is interesting to observe that, when $\theta\rightarrow0$, all the probabilities converge to $\frac{1}{2}$, as it is expected to happen: in fact, in this case, $H_\theta$ becomes the Hamiltonian of the standard quantum harmonic oscillator.

\vspace{2mm}

The same conclusion can be deduced if we compute the three probability transition functions in (\ref{27}) taking $\Phi_0=\varphi_0^{(\theta)}+\varphi_1^{(\theta)}$, and $\Phi_f=\Psi_0^{(\theta)}+\Psi_1^{(\theta)}$. The main difference, in this case, is that an explicit dependence on time appears. In fact we get:
$$
P_{\Phi_0\rightarrow\Phi_f}(t)=\frac{2\cos^3(2\theta)(1+\cos(\omega_\theta)t)}{(1+\cos(2\theta))^2},
$$
and
$$
P_{\Phi_0\rightarrow\Phi_f}^\Psi(t)=P_{\Phi_0\rightarrow\Phi_f}^\varphi(t)=
\frac{\cos^{3/2}(4\theta)(1+\cos^2(2\theta)+2\cos(2\theta)\cos(\omega_\theta t))}{2(1+\cos(4\theta))\cos^3(2\theta)}.
$$
Except for $P_{\Phi_0\rightarrow\Phi_f}(t)$, which is defined for $\theta\in I$, $P_{\Phi_0\rightarrow\Phi_f}^\Psi(t)$ and $P_{\Phi_0\rightarrow\Phi_f}^\varphi(t)$ again make sense only if $\theta\in I_1$. Then the conclusion is the following: when $\Sc$ is driven by an Hamiltonian $H$ which is not self-adjoint (and has real eigenvalues) it seems more convenient not to change scalar product, looking for some other scalar products which makes of $H$ a self-adjoint operator. This is because, if we do that, we may need to restrict the original range of the parameters of the model modifying, in fact, the original model.

\subsection{Extended Landau levels}\label{sectLL}

A similar conclusion can be deduced by considering this third model, whose
 main ingredients are the operators defined in \cite{abg} as follows:
$$
 \left\{
    \begin{array}{ll}
A_1=\frac{1}{\sqrt{2}}\left(\partial_x-i\partial_y+\frac{x}{2}(1+2k_2)-\frac{iy}{2}(1-2k_1)\right),\\
B_1=\frac{1}{\sqrt{2}}\left(-\partial_x-i\partial_y+\frac{x}{2}(1-2k_2)+\frac{iy}{2}(1+2k_1)\right),\\
A_2=\frac{1}{\sqrt{2}}\left(-i\partial_x+\partial_y-\frac{ix}{2}(1+2k_2)+\frac{y}{2}(1-2k_1)\right),\\
B_2=\frac{1}{\sqrt{2}}\left(-i\partial_x-\partial_y+\frac{ix}{2}(1-2k_2)+\frac{y}{2}(1+2k_1)\right),\\
      \end{array}
        \right.$$
        where $k_j\in\left]-\frac{1}{2},\frac{1}{2}\right[$. They satisfy the two-dimensional pseudo-bosonic commutation rules $[A_j,B_k]=\1\delta_{j,k}$. The vacua of $A_j$ and $B_j^\dagger$ are
        $$
 \left\{
    \begin{array}{ll}
\varphi_{0,0}(x,y)=N_\varphi\,\exp\left\{-\frac{x^2}{4}(1+2k_2)-\frac{y^2}{4}(1-2k_1)\right\}\\
\Psi_{0,0}(x,y)=N_\Psi\,\exp\left\{-\frac{x^2}{4}(1-2k_2)-\frac{y^2}{4}(1+2k_1)\right\},\\
\end{array}\right.
$$ where $N_\varphi$ and $N_\Psi$ are normalization
constants which are chosen in such a way that
$\left<\varphi_{0,0},\Psi_{0,0}\right>=1$. We fix them as $N_\varphi=N_\Psi=\frac{1}{\sqrt{2\pi}}$. Of course, since $k_1$ and
$k_2$ are such that $-\frac{1}{2}<k_j<\frac{1}{2}$, $j=1,2$, both these functions are square integrable.

As in \cite{abg} we define the
functions $$
\varphi_{n,l}(x,y)=\frac{B_1^n\,B_2^l}{\sqrt{n!\,l!}}\,\varphi_{0,0}(x,y),
\quad\mbox{ and }\quad
\Psi_{n,l}(x,y)=\frac{(A_1^\dagger)^n\,(A_2^\dagger)^l}{\sqrt{n!\,l!}}\,\Psi_{0,0}(x,y),
$$ where $n,l=0,1,2,3,\ldots$, and the biorthogonal sets
$\F_\Psi=\{\Psi_{n,l}(x,y),\,n,l\geq0\}$ and
$\F_\varphi=\{\varphi_{n,l}(x,y),\,n,l\geq0\}$. Introducing further $h_1=B_1A_1-\frac{1}{2}\1$ and $h_2=B_2A_2-\frac{1}{2}\1$, it is clear that $[h_1,h_2]=0$, and that
$$
h_1\varphi_{n,l}=\left(n-\frac{1}{2}\right)\varphi_{n,l}, \quad
h_2\,\varphi_{n,l}=\left(l-\frac{1}{2}\right)\varphi_{n,l},
$$
and
$$
h_1^\dagger\Psi_{n,l}=\left(n-\frac{1}{2}\right)\Psi_{n,l}, \quad
h_2^\dagger\Psi_{n,l}=\left(l-\frac{1}{2}\right)\Psi_{n,l}.
$$
Furthermore, defining $S_\varphi=e^{-x^2k_2+y^2k_1}$ and
$S_\Psi=S_\varphi^{-1}=e^{x^2k_2-y^2k_1}$, one can check that, for instance, $S_\Psi\varphi_{n,l}=\Psi_{n,l}$, for all $n$ and $l$: in this case, the metric operator is a simple multiplication operator. Also, it is clearly unbounded with unbounded inverse.

\vspace{2mm}

{\bf Remark:--} The particular case $k_1=k_2=0$ returns the {\em standard} Landau levels, \cite{abg}. This can be understood already from what we have discussed here: in this case, in fact, $S_\varphi=S_\Psi=\1$, and the sets $\F_\Psi$ and $\F_\varphi$ collapse to a single set of o.n. functions, complete in $\Lc^2(\Bbb R^2)$.

\vspace{2mm}

Let now assume that the system evolves according to a very simple Hamiltonian: $H=h_1+\frac{1}{2}\1=B_1A_1$, and that it is prepared in the state $\Phi_0=\varphi_{0,0}+\varphi_{1,0}+\varphi_{0,1}$. We want to compute the transition probabilities, as introduced in (\ref{27}), to find $\Sc$ in the final state $\Phi_f=\Psi_{0,0}+\Psi_{1,0}$.

After some lengthy computations we find that
$$P_{\Phi_0\rightarrow\Phi_f}(t)=2(1+\cos(t))p_{k_1,k_2},$$
where
$$
p_{k_1,k_2}:=\frac{\sqrt{(1-4k_1^2)^3(1-4k_2^2)^3}}{(2+3k_1-3k_2-4k_1k_2)(3+4k_1-4k_2-4k_1k_2)},
$$
while the expression for $P_{\Phi_0\rightarrow\Phi_f}^\Psi(t)$ looks more complicated, since it is not possible to separate the dependence on time and on $k_j$ in the final formula:
$$
P_{\Phi_0\rightarrow\Phi_f}^\Psi(t)=\left(1+\left(\frac{1+k_1-k_2}{(1+2k_1)(1-2k_2)}\right)^2+2\cos(t)\frac{1+k_1-k_2}{(1+2k_1)(1-2k_2)}
\right)g_{k_1,k_2},
$$
where
$$
g_{k_1,k_2}:=\frac{(1+4k_1)(1-4k_2)}{6(1+2k_1)(1-2k_2)(1+3k_1-3k_2-8k_1k_2)}.
$$
A similar formula could also be deduced for $P_{\Phi_0\rightarrow\Phi_f}^\varphi(t)$. Notice that we get $P_{\Phi_0\rightarrow\Phi_f}(t) = P_{\Phi_0\rightarrow\Phi_f}^\Psi(t)=\frac{1}{3}(1+\cos(t))$ if we go back to the ordinary Landau levels, i.e. if we take $k_1=k_2=0$.
 It should be emphasized here that $P_{\Phi_0\rightarrow\Phi_f}^\varphi(t)$ can be found only under the additional requirement that $k_j$ takes value only in $\left]-\frac{1}{4},\frac{1}{4}\right[$ rather than in the original, larger set. This is very close to what we have seen for the Swanson model, and again the suggestion is that, if we don't like this kind of additional restrictions on the parameters of the model, we have to choose, among the possibilities given in (\ref{27}), the {\em original one}, i.e. $P_{\Phi_0\rightarrow\Phi_f}(t)$, working with the scalar product $\left<.,.\right>$: this is something new with respect with what was found in \cite{bagAoP}, and with what is quite often discussed in the literature, since is a phenomenon which can only be seen in an infinite dimensional Hilbert space.

\section{Conclusions}\label{sectconl}

After a general discussion which extends to an infinite dimensional framework what originally proposed for  the dynamical problem generated by a non self-adjoint Hamiltonian acting on finite dimensional Hilbert space, we have deduced some consequences of our choices in the computation of several, inequivalent, transition probabilities. With the help of three simple examples we have seen that, in order to keep unchanged the original features of the model under analysis, and in particular the ranges of the parameters defining the model, the only possible choice of the transition probability is $P_{\Phi_0\rightarrow\Phi_f}(t)$. This, in turn, suggests that the only realistic scalar product is the original one, $\left<.,.\right>$, while $\left<.,.\right>_\Psi$ and $\left<.,.\right>_\varphi$ should be understood only as auxiliary useful tools in the analysis of the model, but not really essential or having any deep physical interpretation.

Of course, this is not really so if we admit the possibility of changing the model {\em on the way}, i.e. to further restrict the values allowed for the parameters to ensure, as in this paper, the square-integrability of some relevant functions to respect to some particular metric. In this case, in fact, all the functions defined in (\ref{27}) are on the same footing, at least for the models considered in Section \ref{sectexamples}, and can only be discriminated by some experiments.

We want to stress once again that these features were completely hidden in our previous analysis, \cite{bagAoP}, and in many of the papers existing in the literature, since are intrinsically related to the explicit appearance of infinite dimensional Hilbert spaces.

\section*{Acknowledgements}
The author would like to acknowledge  support from the
   Universit\`a di Palermo and from Gnfm.

\end{document}